# Anomalous Thermal Hall Effect in an Insulating van der Waals Magnet VI$_3$

Heda Zhang[1,+], Chunqiang Xu[1,2,+], Caitlin Carnahan[3,+], Milos Sretenovic[1], Nishchay Suri[3], Di Xiao[4,5], and Xianglin Ke[1]

[1]*Department of Physics and Astronomy, Michigan State University, East Lansing, Michigan 48824-2320, USA*

[2]*School of Physics Southeast University, Nanjing 211189, China*

[3]*Department of Physics, Carnegie Mellon University, Pittsburgh, Pennsylvania 15213, USA*

[4]*Department of Materials Science and Engineering, University of Washington, Seattle, Washington 98195, USA*

[5]*Department of Physics, University of Washington, Seattle, Washington 98195, USA*

[+] These authors contributed equally to this work.

Two-dimensional (2D) van der Waals (vdW) magnets have been a fertile playground for the discovery and exploration of physical phenomena and new physics. In this Letter, we report the observation of an anomalous thermal Hall effect (THE) with $\kappa_{xy} \sim 1 \times 10^{-2}\ W\ K^{-1}\ m^{-1}$ in an insulating van der Waals ferromagnet VI$_3$. The thermal Hall signal persists in the absence of an external magnetic field and flips sign upon the switching of the magnetization. In combination with theoretical calculations, we show that VI$_3$ exhibits a dual nature of the THE, i.e., dominated by topological magnons hosted by the ferromagnetic honeycomb lattice at higher temperatures and by phonons induced by the magnon-phonon coupling at lower temperatures. Our results not only position VI$_3$ as the first ferromagnetic system to investigate both anomalous magnon and phonon THEs, but also render it as a potential platform for spintronics/magnonics applications.

Recently, the thermal Hall effect (THE), i.e., the generation of a transverse heat current in the presence of a longitudinal temperature gradient, has been gaining increasing attention as a powerful tool to probe charge neutral quasiparticles in insulating quantum materials [1-14]. In particular, the THE has been predicted in magnets possessing nontrivial spin excitations, e.g., magnons [15-22]. Distinct from the ordinary charge Hall effect in which charge carries are deflected by the Lorentz force, the transverse flow of magnon current [Fig. 1(a)] is driven by the Berry curvature resulting from the nontrivial topology of the magnon bands. The Berry curvature of magnons usually arises from the Dzyaloshinskii-Moriya interaction (DMI), which depends sensitively on the lattice geometry [8]. Prototypical lattice structures capable of exhibiting the magnon THE include pyrochlore lattice in 3D [1] and kagome lattice in 2D [7], in both of which a nearest neighbor DMI is responsible for the appearance of the Berry curvature. Nevertheless, magnon THE has only been reported in a few ordered magnets to date and the observed thermal Hall conductivity in these materials is rather weak, in the order of $1 \times 10^{-3} W\ K^{-1} m^{-1}$ [1,7,8].

The recent emergence of 2D van der Waals (vdW) magnets [23] provide a new platform in pursuit of a large THE. Specifically, in many of these layered materials, the magnetic ions form a honeycomb lattice in each layer. It has been predicted that magnon bands of such a system could possess non-zero Berry curvature due to a next-nearest-neighbor DMI and can be characterized by non-trivial topological invariants [24]. As a result, the THE of magnons is expected [21,25]. The DMI in these materials can be substantial [26-28], and the DMI - as opposed to Kitaev interactions [29] - has been demonstrated to be the microscopic origin of the Dirac gap revealed in inelastic neutron scattering studies [28]. However, direct experimental evidence of the magnon THE in 2D vdW magnets is still lacking.

In this Letter, we report the observation of an anomalous THE in a vdW ferromagnetic insulator VI$_3$ ($T_c$ = 50 K). The thermal Hall conductivity reaches a magnitude of $\kappa_{xy} \sim 1 \times 10^{-2}$ $W\,K^{-1}m^{-1}$ in a wide temperature range (10 K to 45 K) and the maximum thermal Hall angle ($\theta = \kappa_{xy}/\kappa_{xx}$) is about $4.3 \times 10^{-3}$. We find that the thermal Hall conductivity tracks nicely the magnetization hysteresis loop and persists at zero magnetic field. In conjunction with theoretical calculations, we show that the observed thermal Hall signal at higher temperatures is mainly determined by the non-trivial magnon band topology while, interestingly, at lower temperatures the THE is driven by the magnon-phonon coupling and is of phononic nature.

VI$_3$ crystalizes in the trigonal space group $R\bar{3}$ (No. 148) at room temperature with lattice constants $a = b = 6.8325$ Å, $c = 19.6776$ Å and crystalline angles of $\alpha = \beta = 90°$, $\gamma = 120°$. The vanadium atoms form a honeycomb structure in the *ab*-plane, as illustrated in Figure 1(b). VI$_3$ is a highly resistive semiconductor with a bandgap of ~ 0.67 eV [30,31]. It undergoes a subtle structural phase transitions at $T_s \sim$ 77 K, followed by a paramagnetic-ferromagnetic phase transition at $T_c \sim$ 50 K [30-32]. Figure 2(a) shows the temperature dependence of magnetic susceptibility $\chi_c(T)$ of a VI$_3$ single crystal measured under the field-cooled condition. A magnetic field of 0.1 T is applied along the *c*-axis. It is clearly seen that the sample exhibits a ferromagnetic phase transition at $T_c$ = 50 K as determined by the peak in $d\chi/dT$ (black curve, arbitrary unit). The magnetic easy axis is found to be along the *c*-axis, which is supported by the isothermal magnetization measurements shown in Supplementary Figure S2 [33]. These features are consistent with previous reports [30-32,34], affirming the high quality of our sample.

Figure 2(b) presents the temperature dependence of longitudinal thermal conductivity ($\kappa_{xx}$). As illustrated in the experimental setup shown in Figure 1(a), the heat current is applied in the *ab*-plane. In the absence of magnetic field, there are two anomalies clearly seen in $\kappa_{xx}$ at $T_s$ =

77 K and at $T_c$ = 50 K, respectively. The characteristic broad peak in $\kappa_{xx}$ around 16 K is the result of the competition between the Umklapp phonon scattering and the phonon scattering by defects that dominate in different temperature regions. Interestingly, there is a slight dip in $\kappa_{xx}$ of VI$_3$ below $T_c$ in the absence of magnetic field. As will be discussed later, this feature suggests that there is strong magnon-phonon scattering in VI$_3$ which leads to a reduction of thermal conduction by phonons. Such a reduction even surpasses the thermal conductivity contributed by magnon carriers, resulting in the decrease of total $\kappa_{xx}$ seen in Figure 2(b).

The isothermal magnetization ($M$), $\kappa_{xx}$, and $\kappa_{xy}$ of VI$_3$ as a function of the applied magnetic field along the *c*-axis are presented in Figure 3. As illustrated in Figure 1(a), the sign convention of $\kappa_{xx}$ and $\kappa_{xy}$ is determined by $\Delta T_{xx} = T_1 - T_2$ and $\Delta T_{yx} = T_3 - T_2$, respectively. In the upper and middle panel of Figure 3(a), we plot $M$ and $\kappa_{xx}$ as a function of magnetic field measured at $T$ = 15 K, both of which exhibit typical hysteresis behavior for a ferromagnet with easy-axis magnetic anisotropy: a bow-tie shaped hysteresis loop in $\kappa_{xx}$ is observed with a sharp increase occurring at the coercive field. A similar feature in $\Delta\kappa_{xx}(H)$ has been observed at other temperatures, as shown in Supplementary Figure S3 [33]. As mentioned above, the total thermal conductivity in an insulator can have magnon and phonon contributions, $\kappa^{tot} = \kappa^{mag} + \kappa^{ph}$. Upon increasing the magnetic field strength, the magnon population is suppressed and $\kappa^{mag}$ decreases. On the other hand, in a system where phonons are strongly scattered by magnons, the suppression of magnons can lead to longer mean free path for phonons, thereby increasing $\kappa^{ph}$. This is also seen from the comparison of $\kappa_{xx}(T)$ measured at 0 T and 7 T shown in Figure 2(b). In VI$_3$, the total longitudinal thermal conductivity increases with magnetic field, which suggests that the longitudinal heat current is mainly carried by phonons and the phonons are strongly scattered by magnons.

The lower panel of Figure 3(a) shows the field dependent $\kappa_{xy}$ measured at $T$ = 15 K. For reference, two vertical dashed lines at the coercive field are plotted across the three panels in Figure 3 (a). It is clearly seen that the $\kappa_{xy}(H)$ curve tracks the $M(H)$ curve very well and $\kappa_{xy}$ saturates right above the coercive field. Remarkably, the magnitude of $\kappa_{xy}$ reaches ~ 0.01 $W\ K^{-1}m^{-1}$, nearly an order of magnitude larger than the thermal Hall signal observed in other ordered magnets [1,7,8]. Similar behavior is also observed at higher temperatures up to $T_c$, as shown in Figure 3(b) and Figure S4 [33]. At lower temperatures, due to the strong mixing between the longitudinal and transverse components, an antisymmetrization procedure needs to be carried out [7,35], which allows us to extract saturated value of $\kappa_{xy}$ (see Figure S5 and S6 in the Supplemental Material [33]). The thermal Hall angle, which is defined as $\theta = \kappa_{xy}/\kappa_{xx}$, reaches a maximum value of ~ $4.3 \times 10^{-3}$ at $T$ = 44 K. With a stronger magnetic field applied, as shown in Fig. S7, the magnitude of $\kappa_{xy}$ decreases while $\kappa_{xx}$ increases and the magnetization $M$ remains nearly saturated. Thus, the $\kappa_{xy}(H)$ curves suggest that the mechanism responsible for THE in VI$_3$ should be closely related to its spin degrees of freedom, i.e., magnons, since the suppression of magnon population at high field leads to a reduction in $\kappa_{xy}$ and an increase in $\kappa_{xx}$ due to the suppression of phonon scattering by magnons. Furthermore, the thermal Hall signal persists in the absence of an external magnetic field with $\kappa_{xy}$ being nearly the same as the saturated value, indicating an anomalous THE, and $\kappa_{xy}$ flips sign upon the switching of the magnetization. These merits, together with its quasi-2D nature, demonstrate that VI$_3$ could be a promising platform for future spintronics/magnonics applications [36-40].

To understand the physical origin of the anomalous THE observed in VI$_3$, we examine the corresponding 2D Heisenberg model of spins $\boldsymbol{S}_i$ on a honeycomb lattice. Since VI$_3$ is a 2D vdW ferromagnet with a large spacing between layers along the c-axis and a much weaker interlayer magnetic exchange interaction than the intralayer exchange interactions, the magnon band structure is mainly determined by the intralayer interactions. This has been elaborated in recent inelastic neutron scattering studies of sister compounds $CrI_3$ and $CrBr_3$ [26-28]. It has previously been predicted that a second-nearest neighbor DMI should be allowed in such a system [24], and this interaction gives rise to topologically-driven response functions in honeycomb ferromagnets [21,25,41]. The Hamiltonian is given by

$$H = J_1 \sum_{\langle i,j \rangle} \mathbf{S}_i \cdot \mathbf{S}_j + J_2 \sum_{\langle\langle i,j \rangle\rangle} \mathbf{S}_i \cdot \mathbf{S}_j + J_3 \sum_{\langle\langle\langle i,j \rangle\rangle\rangle} \mathbf{S}_i \cdot \mathbf{S}_j + \sum_{\langle\langle i,j \rangle\rangle} \mathbf{D}_{ij} \cdot \mathbf{S}_i \times \mathbf{S}_j - K \sum_i S_{i,z}^2$$

$$- \sum_i B S_{i,z} \quad (1)$$

where the symmetric exchange parameters $J_1$, $J_2$, and $J_3$ describe interactions between nearest, second-nearest, and third-nearest neighbors, respectively, such that $J < 0$ leads to ferromagnetic coupling. The Dzyaloshinskii-Moriya vector $\mathbf{D}_{ij} = \pm D\hat{z}$ is directed out-of-plane and easy-axis anisotropy and Zeeman coupling are parameterized by $K$ and $B$ respectively (additional details and parameter values can be found in the Supplemental Materials [33,42]). The magnon band structure is shown in Figure 4(a). The blue and red lines are the magnon dispersion calculated along $\Gamma - K$ with $D = 0$ meV. As expected, there is symmetry-protected linear band touching at the K-points, and the magnon bands remain topologically trivial [24]. The surface plots represent the spin wave dispersion with $D = 0.2$ meV included. We can see that a non-zero DM interaction opens a gap between the magnon bands at each K-point [24,41]. Such a feature is akin to that found in the Haldane model [43], where exchange interactions ($J_s$) play the role of hopping integrals and DM interaction plays the role of spin-orbit coupling. Since the two magnon bands are fully gapped in the reciprocal space, we can calculate the Berry curvature and the corresponding Chern number for each band.

The Berry curvature projection along the $c$-axis ($|\Omega_z|$) of the lower magnon band is presented in Figure 4(b). We see that non-zero $\Omega_z$ is found along the K − M − K lines in the reciprocal space. By integrating $\Omega_z$ over the first Brillouin zone, the Chern number for the upper and lower magnon bands are found to be C = +1 and C = -1, respectively. Due to the bulk-boundary correspondence, edge states are anticipated to connect these two bands with different topological indices.

To further verify that the observed $\kappa_{xy}$ in VI$_3$ is indeed driven by magnons, we have calculated the temperature dependence of the thermal Hall conductivity $\kappa_{xy}$ of the model introduced in Eq. 1 and compare it with the experimental results. It is standard practice to deduce the 3D thermal conductivity from the calculated 2D "sheet conductivity" through division by the interlayer distance of the material. To cover the entire temperature range, we used a mean-field treatment of the magnetization in the Holstein–Primakoff transformation [44]. The details of the calculation are described in the Supplemental Materials [33]. The calculated $\kappa_{xy}$ at a magnetic field of $B = 1.0$ T as a function of temperature is shown in Figure 4(c) for various strength $D$ of the DMI, ranging from $D/J_1 \approx 0.0036$ to $D/J_1 \approx 0.67$ [for reference, $D/J_1 \approx 0.089$ in CrI$_3$ [45] and $\approx 0.15$ in CrBr$_3$ [27]]. There are a couple of features worth pointing out. First, the calculated $\kappa_{xy}$ is strongly dependent on the magnitude of $D$, and a non-zero $D$ is crucial for generating the thermal Hall signal. Second, for reasonable values of D, the maximum of the calculated $\kappa_{xy}$ agrees well with the magnitude of the experimental results.

In Figure 4(d) we plot the temperature dependence of the saturated magnitude of experimental $\kappa_{xy}$, together with the calculated $\kappa_{xy}$ for $D = 0.2$ meV ($D/J_1 \approx .073$). The calculated thermal Hall conductivity (in blue) and the experimental results (in red) show a qualitative agreement particularly at higher temperature. In the vicinity of $T_c$, the difference between the calculated and experimental $\kappa_{xy}$ values are expected since the mean-field approach cannot accurately capture fluctuation phenomena near the phase transition. Below 30 K, the experimental $\kappa_{xy}(T)$ forms a shoulder-like feature and decreases more slowly than the calculated value when approaching $T = 0$ K. Such a feature has been observed in two other VI$_3$ samples (S1 and S3) measured [Figure S8 [33]]. Figure 4(e) plots the temperature dependence of $\Delta\kappa_{xy}$, the difference between the experimental $\kappa_{xy}$ values and the calculated $\kappa_{xy}$ shown in Fig. 4(d). For comparison, the temperature dependence of the longitudinal thermal conductivity $\kappa_{xx}$ measured at zero field, which is dominated by phonons as heat carriers, is also presented in Fig. 4(e). We can see that $\Delta\kappa_{xy}$ and $\kappa_{xx}$ peak at nearly the same temperature (~ 14 K), which suggests the dominant phononic origin of the THE at lower temperatures. The high-temperature peak of $\Delta\kappa_{xy}$ around 50 K (i.e., T$_c$) is an artificial feature, because the calculated $\kappa_{xy}$ based on the mean-field approach cannot accurately capture fluctuation phenomena near the phase transition. An intriguing question arises: How do phonons play a role?

First, we can exclude the possibility that the THE observed in VI$_3$ is purely driven by phonons that directly couple to the magnetic field as in the case of nonmagnetic SrTiO$_3$ [13]. In this scenario, the THE should vanish at zero field and increase with magnetic field, but in VI$_3$ we observe a large thermal Hall signal at zero field, i.e., anomalous THE. This suggests that the thermal Hall signal observed in VI$_3$ is associated with the magnetism. Second, we can also exclude the possibility that the observed thermal Hall signal in VI$_3$ is mainly driven by the phonon THE induced by the internal magnetic field of magnetization. This is evident by the comparison of M(H) and $\kappa_{xy}(H)$ measured in the high field region as shown in Fig. S7, where we can see that M saturates above 0.6 T while $\kappa_{xy}$ shows a continuous decrease at higher field. Provided that a phonon THE induced by the internal magnetic field of magnetization dominates at low temperature, one would expect $\kappa_{xy}$ to be directly proportional to magnetization. The discrepancy in the trend of *M*(*H*) and $\kappa_{xy}(H)$ suggests that magnetic excitations instead of static magnetization be directly related to the THE.

Instead, a likely potential source of the low-temperature discrepancy between the calculated magnon-mediated $\kappa_{xy}$ and the measured values is potential magnon-phonon coupling [46-50]. To explore this possibility, we have studied a simple proof-of-concept model in which the magnon-phonon interaction is incorporated through magnetoelastic coupling (see Supplemental Materials[33] for details). We find that the inclusion of magnon-phonon coupling introduces non-zero Berry curvature at relatively low energies near the Γ-point where anticrossings between the phonon and magnon bands develop [Fig. S9(a-d)], which indeed results in a rapid increase of $\kappa_{xy}$ when the temperatures increases from 0 K, as shown by the green dots in Fig. 4(d) and in Fig. S9(e) [33]. While we note that the resultant $\kappa_{xy}$ does not completely account for the larger thermal Hall signal observed at low temperature, we would like to point out that the calculation utilizes a fairly simple toy model in which only the out-of-plane phonon mode (i.e., ZA mode) of the vanadium atoms are considered and certain values of magnon-phonon coupling constants are assumed. More sophisticated calculations considering various phonon modes (including ZA, transverse acoustic and longitudinal acoustic modes) of all atoms and their couplings to the magnon modes are desirable. A full understanding of the role of magnon-phonon coupling will also require detailed studies of the spin and lattice dynamics in VI$_3$ via inelastic neutron/resonant x-ray scattering. Nevertheless, it is worth pointing out that when *D* = 0 the calculated $\kappa_{xy}$ due to magnon-phonon interaction alone is much smaller [Figure S9(f) [33]] and has a qualitatively different temperature dependence than that observed experimentally. Interestingly, the calculated THE induced by magnon-phonon coupling shows a peak at lower temperature, as plotted in Fig. S9(f), which is consistent with the $\Delta\kappa_{xy}$ presented in Fig. 4(e). This suggests the dominant phononic origin of the THE observed at lower temperatures, which is mainly induced by the magnon-phonon coupling. In the presence of a strong magnetic field, the magnetic field shifts the magnon band upwards in energy, which in turn moves the anticrossing point upwards, leading to a reduction of $\kappa_{xy}$, as presented in Fig. S10. This is consistent with the experimental observation shown in Fig. S7.

Figure 4(f) presents the anomalous thermal Hall angle $\theta = \frac{\kappa_{xy}}{\kappa_{xx}}$ as a function of temperature. At lower temperatures (below 20 K) the phonon THE dominates with $\theta$ within [0.17% 0.25%]; with increasing magnon population at higher temperatures, the magnon THE contribution increases and dominates with a much larger $\theta \approx 0.43\%$ at 44 K. Our work not only bridges the gap between the theoretical and experimental studies in the emerging field of topological magnons, but also points to a promising candidate material for utilizing the unique character of topological magnons in 2D spintronics/magnonics devices [36-40]. For instance, it would be interesting to study the inverse spin Hall effect in this system. Furthermore, the dual nature of THE, i.e., the THE driven by phonons induced by magnon-phonon coupling at lower temperatures and driven by topological magnons at higher temperatures, places $VI_3$ as the first ferromagnetic system to investigate both anomalous magnon and anomalous phonon THEs.

**Acknowledgements**
H.Z., M.S., and X.K. acknowledge the financial support by the U.S. Department of Energy, Office of Science, Office of Basic Energy Sciences, Materials Sciences and Engineering Division under DE-SC0019259. Work at Carnegie Mellon University is supported by AFOSR MURI 2D

MAGIC (FA9550-19-1-0390). C.X. is partially supported by the Start-up funds at Michigan State University.

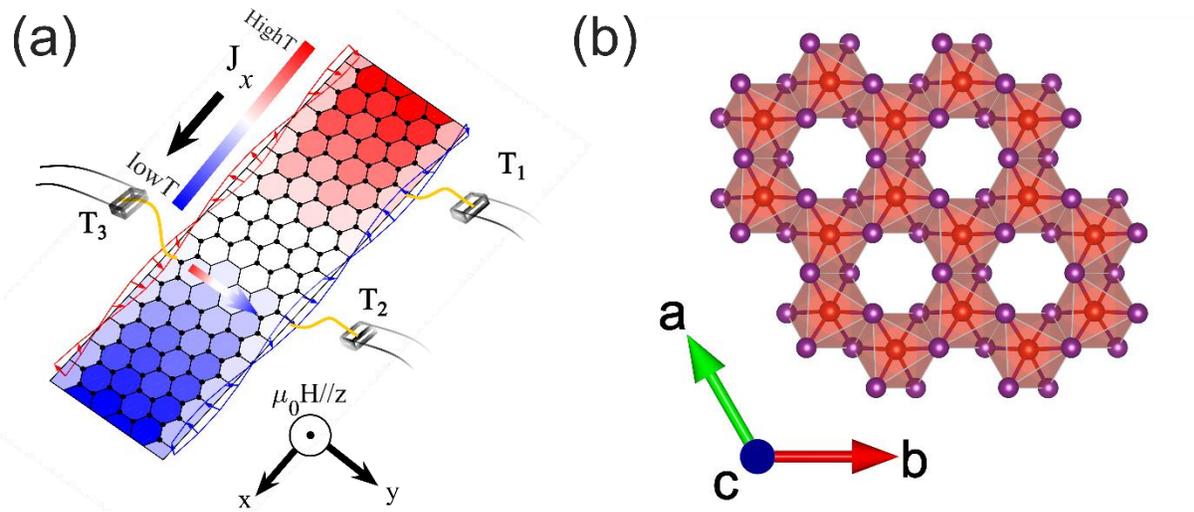

Figure 1. **The experimental set-up and crystal structure of VI$_3$**. (a) A schematic of the experiment set-up. The color indicates temperature profile. A heat current ($J_x$) passes through the sample, and the temperatures at three locations are measured using Cernox sensors. The magnetic field is applied along the out-of-plane direction. (b) The schematic crystal structure of VI$_3$ viewed in the *ab* plane. Vanadium atoms are in red and Iodine atoms are in purple.

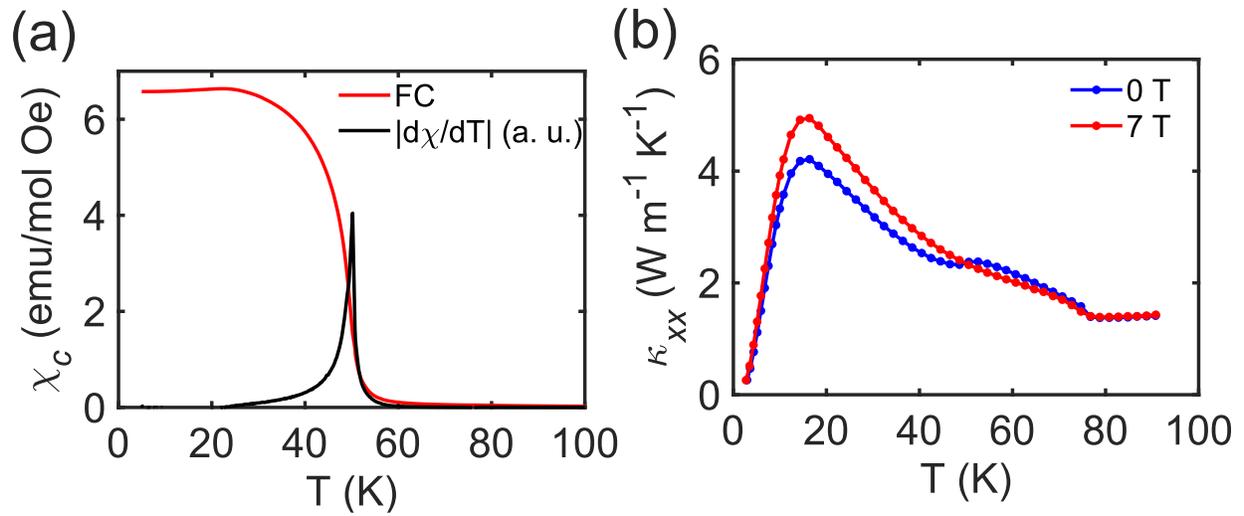

Figure 2. **Temperature dependence of magnetic susceptibility (χ) and longitudinal thermal conductivity ($\kappa_{xx}$) of VI$_3$**. (a) χ(*T*) with 0.1 T magnetic field applied along the *c*-axis. (b) $\kappa_{xx}$(*T*) measured with 0 T and 7 T magnetic field.

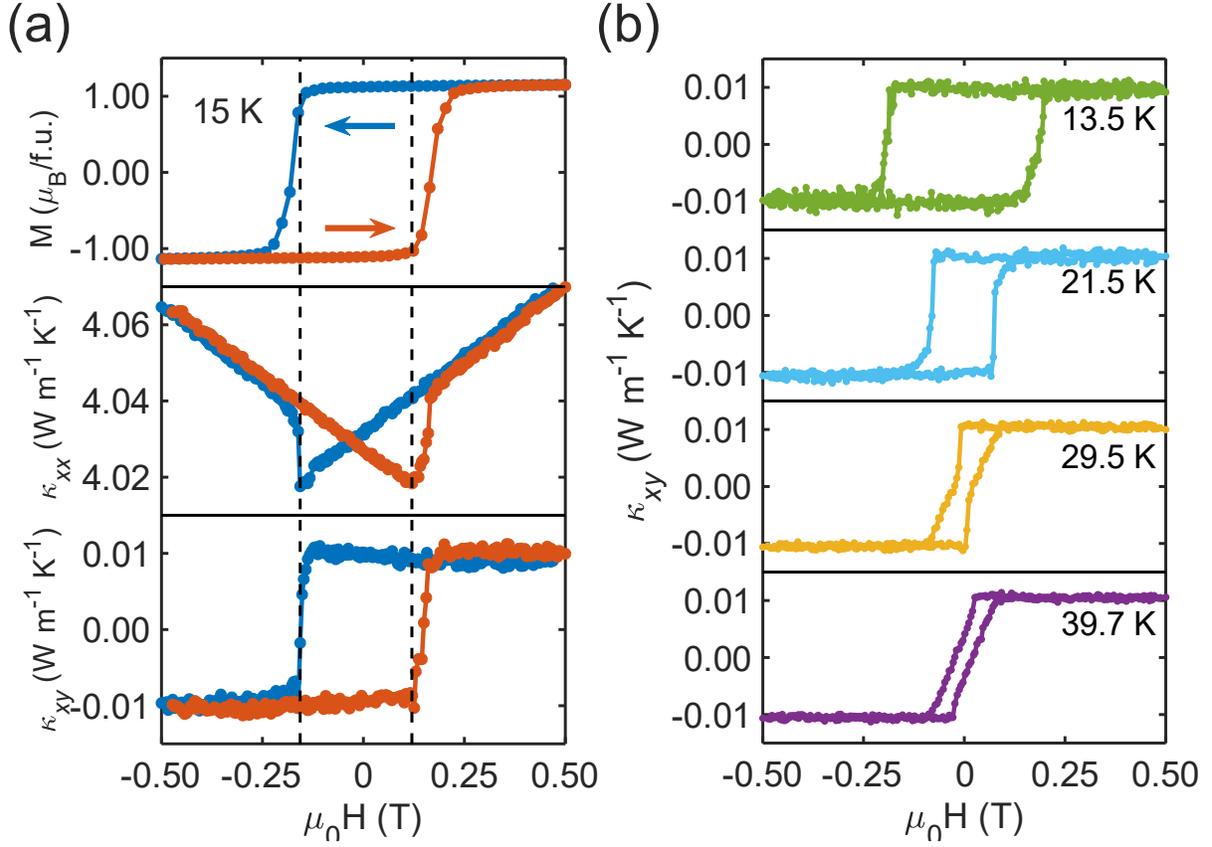

Figure 3. **Magnetic field dependence of magnetization (*M*), longitudinal thermal conductivity ($\kappa_{xx}$) and thermal Hall conductivity ($\kappa_{xy}$) of VI$_3$.** (a) Isothermal $M(H)$, $\kappa_{xx}(H)$, and $\kappa_{xy}(H)$ curves measured at $T = 15$ K. The blue and orange arrows represent the magnetic field sweeping directions. (b) $\kappa_{xy}(H)$ measured at some other selected temperatures. Data of $\kappa_{xy}(H)$ measured below 10 K are shown in Figure S4 & S5.

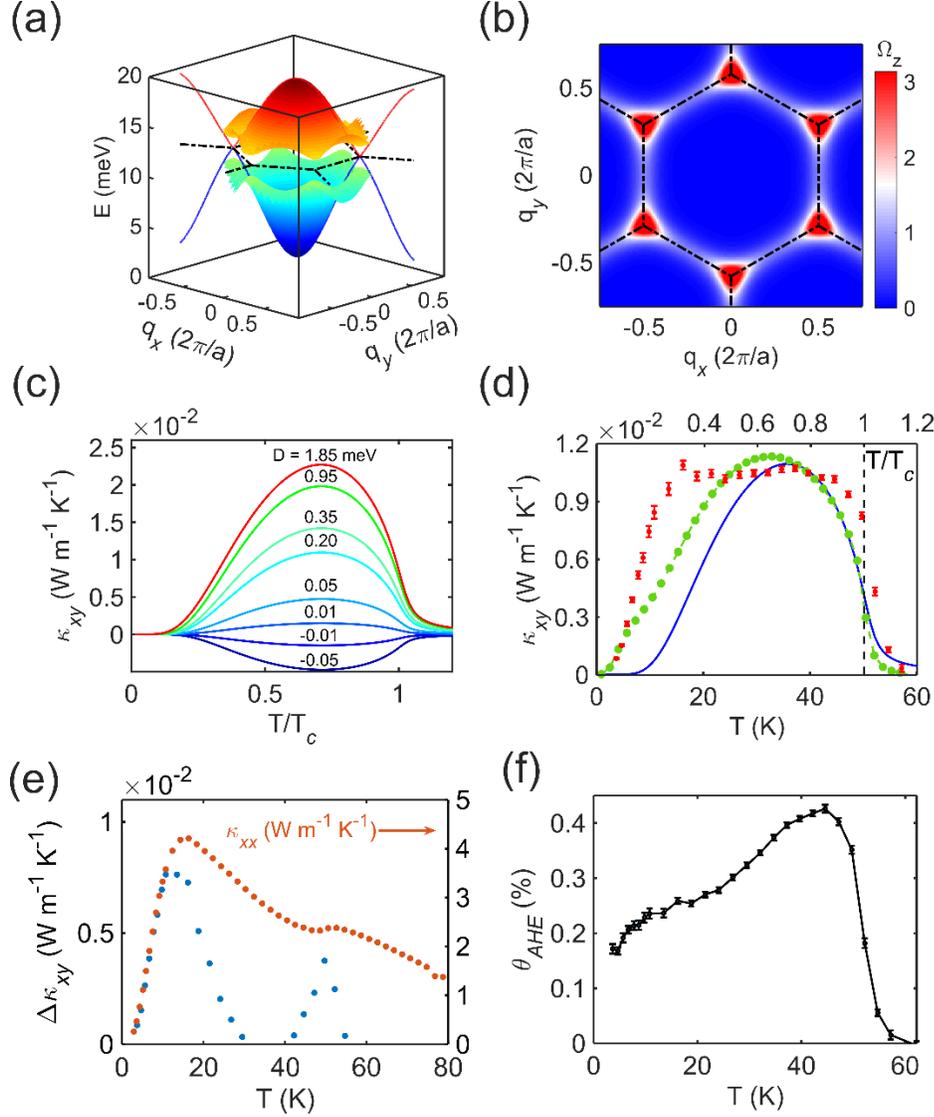

Figure 4. **The calculated magnon bands and the temperature dependence of thermal Hall conductivity ($\kappa_{xy}$) of VI$_3$.** (a) Calculated magnon bands. The black dashed lines illustrate the Brillouin zone boundaries. The blue and red lines show the spin wave dispersions along $\Gamma - K - \Gamma - K - \Gamma$ direction without including the next-nearest DM interaction. The surface plots represent the magnon bands calculated with $D = 0.2$ meV. (b) The calculated Berry curvature ($|\Omega_z|$) of the lower magnon band. (c) The calculated $\kappa_{xy}(T)$ with various $D$ values and $B = 1.0$ T. (d) Experimental values of anomalous $\kappa_{xy}(T)$. The calculated $\kappa_{xy}(T)$ values without ($g = 0$) and with the magnon-phonon coupling ($g = 0.5$ with $g$ defined in the Supplemental Materials) are plotted

in blue curve and in green dots, respectively. The DM interaction ($D = 0.2$ meV) is considered. (e) Temperature dependence of the difference ($\Delta\kappa_{xy}$) between the measured $\kappa_{xy}$ and the calculated $\kappa_{xy}$ driven by topological magnons. $\kappa_{xx}(T)$ is plotted in orange for comparison. (f) Temperature dependence of thermal Hall angle $\theta = \frac{\kappa_{xy}}{\kappa_{xx}}$.